\documentclass[prd,twocolumn,nofootinbib,showpacs,superscriptaddress]{revtex4-1}

\usepackage{amsfonts}
\usepackage{amsmath}
\usepackage{amssymb}
\usepackage{bm}
\usepackage{dcolumn}
\usepackage{graphicx}
\usepackage{graphics}
\usepackage[latin1]{inputenc}
\usepackage{latexsym}
\usepackage{rotating}
\usepackage{hyperref}
\usepackage[all]{hypcap} 
\usepackage{xspace} 
\usepackage[usenames]{color}
\usepackage{mathrsfs}

\usepackage{ulem}
\definecolor {darkgreen}{rgb}{0.2,0.7,0.2}

\newcommand\be{\begin{equation}}
\newcommand\ba{\begin{eqnarray}}
\newcommand\ee{\end{equation}}
\newcommand\ea{\end{eqnarray}}
\newcommand\bw{\begin{widetext}}
\newcommand\ew{\end{widetext}}

\newcommand{\nn}{\nonumber}

\newcommand{\BL}{{\mbox{\tiny BL}}}

\newcommand{\BH}{{\mbox{\tiny BH}}}
\newcommand{\HH}{{\mbox{\tiny H}}}

\newcommand{\NS}{*}

\newcommand{\mrm}{\mathrm}

\begin{document}
\title{An Entropy-Area Law for Neutron Stars Near the Black Hole Threshold}    

\author{Stephon H.~Alexander}
\affiliation{Brown University, Department of Physics, Providence, RI 02912, USA.}

\author{Kent Yagi}
\affiliation{Department of Physics, University of Virginia, Charlottesville, Virginia 22904, USA.}

\author{Nicol\'as Yunes}
\affiliation{eXtreme Gravity Institute, Department of Physics, Montana State University, Bozeman, MT 59717, USA.}

\begin{abstract} 

Neutron stars exhibit a set of universal relations independent of their equation of state that bears semblance to the black hole no hair relations.  Motivated by this, we analytically and numerically explore other relations that connect neutron star and black hole universality.  By analyzing two different measures, we find that certain rescaled entropies possess a nearly universal behavior. We also discover that when the compactness of neutron stars approaches the black hole limit, the rescaled entropy approaches that of a black hole, and the thermodynamic entropy scales with the stellar surface area in an ever more universal way. 

\end{abstract}

\date{\today}
\maketitle
%

\section{Introduction}
\label{sec:intro}

In the context of General Relativity there is a striking semblance to the no-hair relations of black holes~\cite{Geroch:1970cd,hansen} and the universal relations for compact stars~\cite{I-Love-Q-Science,I-Love-Q-PRD,Pappas:2013naa,Stein:2013ofa,Yagi:2014bxa,Yagi:2016bkt}.  The former states that for an isolated, stationary, and uncharged black hole, the only information accessible to an external observer is its mass and its spin.  Likewise, the latter state that a certain universality arises among a compact star's moment of inertia, its tidal deformability (Love number) and its rotation-induced multipole moments, such that these inter-relations are roughly independent of the equation of state.  Given that neutron stars can collapse to form black holes, it is intriguing to ask to what extent these seemingly universal relations may be related to the universality seen in black holes.  In this work, we initiate such an investigation.

One non trivial pathway to explore a connection between compact stars and black holes is to look at their entropy.  From the seminal work by Bekenstein and Hawking~\cite{Bekenstein:1973ur,Bekenstein:1974ax,Hawking:1974sw,Hawking:1974rv}, we know that, when quantum mechanics is included, black holes radiate and have an entropy that surprisingly scales with the event horizon area. Could it be that this is also the case for neutron stars? And if so, is the approach to the black hole limit universal, depending insensitively on the star's equation of state and internal composition? 

At first sight, the answer is seemingly in the negative. Black holes are very cold, with Hawking temperatures smaller than $10^{-8}$ Kelvin, and thus, they have immense entropies. For a typical astrophysical black hole, the Bekenstein-Hawking entropy is greater than $10^{79} k_{B}$, where $k_{B}$ is the Boltzmann constant. On the other hand, isolated neutron stars have surface temperatures of about $10^6$ Kelvin, and entropies of about $10^{63} k_{B}$, as roughly estimated from the ratio of their binding energy to their temperature. We see then clearly a gap of over 16 orders of magnitude between the entropies of these objects.  

Interestingly, though, the approximately universal relations found in neutron stars do not apply to dimensional quantities, but rather to certain rescaled dimensionless combinations. Consider then the dimensionless, \emph{rescaled} entropy, defined as the product of the entropy to the temperature, divided by the mass of the object. For non-rotating and uncharged black holes, this rescaled entropy is exactly equal to $0.5$, if one uses the Bekenstein-Hawking entropy and the Hawking temperature to calculate it. For non-rotating neutron stars, the rescaled entropy is approximately the compactness of the star, which is about $0.2$. In terms of the rescaled entropy, then, neutron stars are much closer to black holes. 

Given this, one may wonder how this rescaled entropy flows from a neutron star sequence to the black hole limit. Is there approximate universality in this flow? This is interesting because one can show that the rescaled entropy is related to the scaling exponent of the entropy-mass or entropy-compactness relation. If the flow is approximately universal, then this scaling exponent is also approximately universal over many orders of magnitude in physics, from scales relevant to neutron stars (nuclear or quantum-chromodynamics scales) to scales relevant to black holes (the Planck scale). Such a result would hint at the existence of an approximate symmetry that becomes sharper at the threshold of gravitational collapse.  
 
We study these ideas here by focusing on two measures of entropy (thermodynamic~\cite{Oppenheim:2001az,Oppenheim:2002kx} and configurational~\cite{Zurek:1985gd,Gleiser:2013mga,Gleiser:2015rwa}) for an equilibrium sequence of cold neutron stars. Our first finding is that both entropy measures, if properly rescaled with the stellar temperature and mass, exhibit an approximate universality that is in fact enhanced as the neutron star sequence approaches the black hole threshold. Perhaps even more intriguing, this rescaled entropy approaches the rescaled entropy of a black hole as the neutron star compactness approaches the black hole limit in an equation-of-state insensitive fashion. When we combine these results, we can also show that the neutron star thermodynamic entropy scales more and more with the stellar surface area (instead of the stellar volume) as its compactness approaches the black hole limit~\cite{Oppenheim:2001az,Oppenheim:2002kx}. 

The results summarized above hint at the existence of critical phenomena in the collapse of neutron stars to black holes. In physics, critical phenomena is related to scaling relations near second-order phase transitions that are thought to be universal, in the sense that they depend only on a few macroscopic properties of the system, like its size or the range of the interaction. In gravity, critical phenomena was found in the 1990s through detail numerical studies of the dynamics of a scalar field at the threshold between dispersal and collapse into a black hole~\cite{Choptuik:1992jv}. The analysis presented here is different from that because it involves an equilibrium sequence of neutron stars, and thus it is not dynamical. Yet it suggests that a dynamical study could be most interesting, and in fact, it could reveal the emergence of an approximate symmetry, like that discussed in~\cite{Yagi:2014qua,Yagi:2015upa}, but now at the threshold of gravitational collapse.

The remainder of this paper presents the details that led us to the conclusions we summarized above. In Sec.~\ref{sec:entropy} we formulate and numerically evaluate both the thermodynamic and configurational entropy of a neutron star.  In Sec.~\ref{sec:SILQ} we discuss the universality of the entropy of neutron stars in various relevant limits. In Sec.~\ref{sec:entropy-area} we compare our results to black hole entropy and in Sec.~\ref{sec:conclusions} we discuss future directions and speculate about the underlying reason behind these surprising connections between black holes and neutron stars.

\section{The entropy of a neutron star}
\label{sec:entropy}

In this section, we define the notions of entropy we will use in this paper: thermodynamic entropy and configurational entropy. With this at hand, we then present the behavior of the entropies as a function of compactness for a variety of well-motivated equations of state, as well as the interrelations between the entropies themselves.    

\subsection{Thermodynamic Entropy} 
\label{subsect:thermo-entropy}

The concept of thermodynamic entropy can be phenomenologically understood as a measure of the organization of the energy of a system. Let us then consider a spherically symmetric star in thermal equilibrium with mass $M$ and radius $R$ and use the first law of thermodynamics 
\be
\label{eq:firstlaw}
dE = T dS_\mrm{th} - p \, dV + \mu \, dN\,,
\ee
with $E$ the energy, $T$ the temperature, $S_\mrm{th}$ the thermodynamic entropy, $V$ the volume, $\mu$ the chemical potential and $N$ the particle number of the star. We can use this equation to solve for the thermodynamic entropy to find 
\be
\label{eq:therm-entropy}
S_\mrm{th} = \frac{4 \pi}{T_{0}} \int_{0}^{R} r^{2} \left(-g_{tt} \; g_{rr}\right)^{1/2} \left(\rho + p - \mu \; n\right) dr\,,
\ee
where $R$ is the stellar radius, $\rho$ is the energy density, $n$ is the number density, and we have used that the volume element for a spherically symmetric metric can be written as $dV = 4 \pi (g_{rr})^{1/2} r^{2} dr$ in a suitably-adapted coordinate system. We have also used that the temperature of a star in internal thermal equilibrium is given by the Tolman relation~\cite{Oppenheim:2002kx}
\be
\label{eq:Tolman}
T(r) = T_{0} \left(-g_{tt}\right)^{-1/2}\,,
\ee
where $T_{0}$ is the temperature of the star as measured at infinity. Notice that the zero temperature limit is delicate in the above expression, as one may naively conclude that it forces the entropy in Eq.~\eqref{eq:therm-entropy} to diverge. In this paper, we study the combination $S_\mrm{th} T_0$, which remains finite even in the zero temperature limit. 

For a system in thermal equilibrium, the chemical potential at any two points inside the star are related to the chemical potential at the surface via
\be
\mu(r) = \mu(R) \left(\frac{g_{tt}(R)}{g_{tt}(r)}\right)^{1/2}\,,
\ee
if there is no net particle flow~\cite{Oppenheim:2001az}. In Appendix~\ref{funky-relation}, we prove that 
\be
\label{eq:funky-relation}
\frac{\mu(R) \, n}{\rho + p} \left(\frac{g_{tt}(R)}{g_{tt}(r)}\right)^{1/2} = c_{\mu} = {\textrm{const.}}\,,
\ee
neglecting temperature corrections in the equation of state. Using this relation, the thermodynamic entropy can be written as
\be
\label{eq:therm-entropy-new}
S_\mrm{th} = \frac{4 \pi}{\bar{T}} \int_{0}^{R} r^{2} \left(-g_{tt} \; g_{rr}\right)^{1/2} \left(\rho + p\right) dr\,,
\ee
where we have defined the renormalized temperature 
\be
\label{eq:Tbar}
\bar{T} \equiv \frac{T_{0}}{1+c_{\mu}}\,.
\ee
%

Let us now plot the thermodynamic entropy for neutron stars using a wide class of realistic equations of state. We first consider neutron stars that can be described by an isotropic, perfect fluid, stress energy tensor with the following equations of state: an $n=0$ (constant density) polytrope, AP4~\cite{APR}, SLy~\cite{SLy}, LS~\cite{LS} with nuclear incompressibility of 220MeV (LS220) and Shen~\cite{Shen1,Shen2}, with the latter two at a temperature of 0.1MeV\footnote{Strictly speaking, these constant temperature equations of state are inconsistent with the Tolman relation in Eq.~\eqref{eq:Tolman}. However, since 0.1MeV is much smaller than the Fermi temperature of a neutron star, we can neglect any thermal radiation and associated dynamics as a first approximation.} with a neutrino-less and beta-equilibrium condition. All of these equations of state support neutron stars with a mass above $2M_\odot$, and thus, they are not yet ruled out by pulsar observations. We also consider neutron stars that can be described by fluids with anisotropic pressure through the simple and phenomenological model of Bowers and Liang~\cite{1974ApJ...188..657B} (the so-called BL model with the strongly-anisotropic limit characterized by an anisotropic parameter $\lambda_{\BL} = -2 \pi$). This class of stars allows for stable solutions with maximum compactness much closer to the black hole limit~\cite{Yagi:2015upa,Yagi:2016ejg}. 

With the equations of state specified, we can now compute the thermodynamic entropy for neutron stars. We start by choosing a central density and numerically solving the differential equations of relativistic structure for a specific equations of state in the list described above. The solution to these equations yields the metric functions $(g_{tt},g_{rr})$ and the thermodynamic variables $(p,\rho)$ as functions of the radial coordinate. The thermodynamic entropy can then be calculated through Eq.~\eqref{eq:therm-entropy-new}, where the radius of the star is given by the radial position at which the pressure vanishes, and the total mass of the star as the value of the enclosed mass at that radial position. We then repeat this calculation over many different values of the central density to find a sequence of neutron stars of varying compactness, storing the values of the radius, total mass, compactness and thermodynamic entropy for each member in the sequence. Finally, we repeat this calculation for a different equation of state, finding sequences of neutron stars for each equation of state listed above. 

\begin{figure}[htb]
\includegraphics[width=0.48\textwidth]{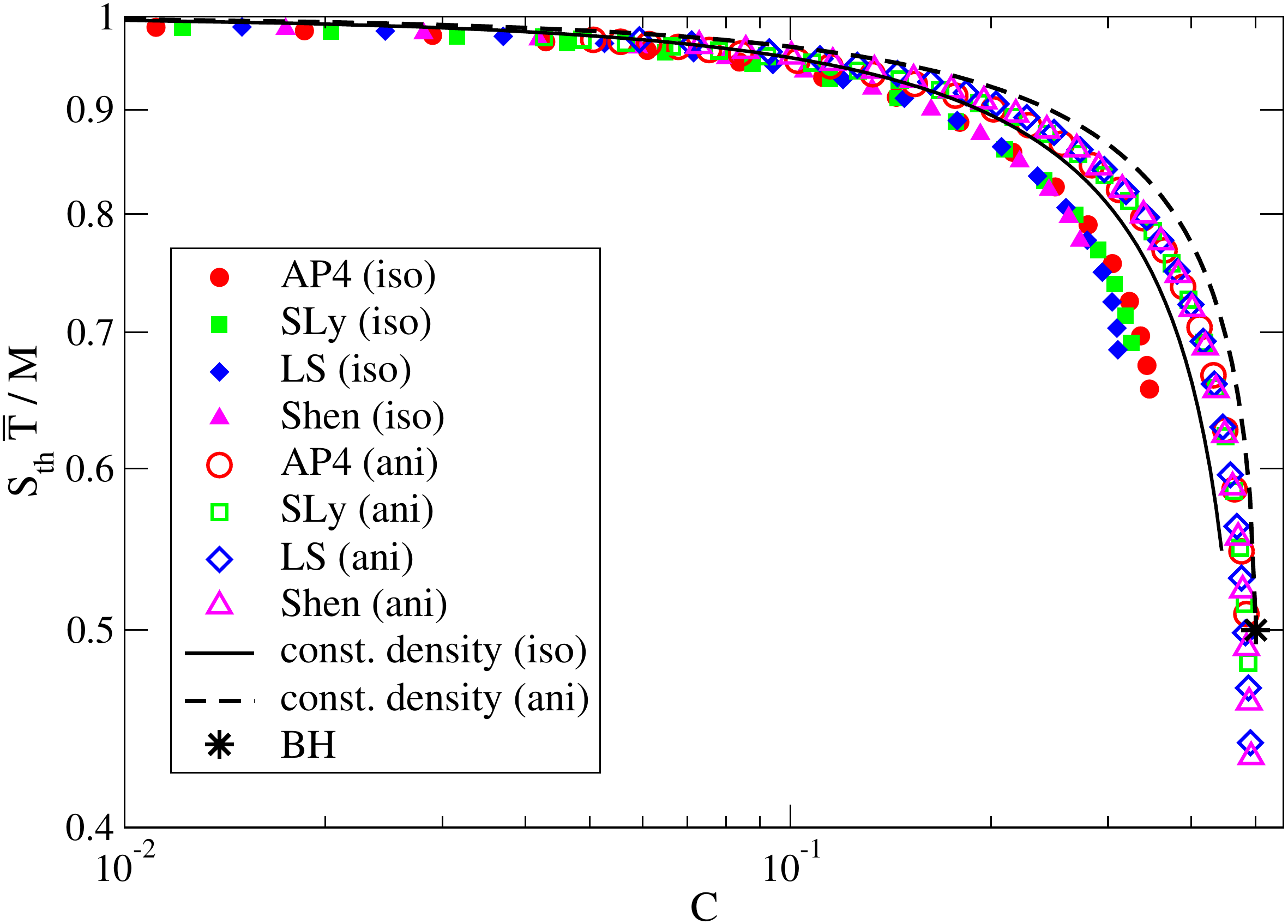}
\caption{Thermodynamic entropy of a sequence of neutron stars of different compactness $C$, scaled by the renormalized temperature and the neutron star mass. The filled symbols correspond to neutron stars with interiors modeled through a perfect fluid stress-energy tensor with isotropic pressure, while the unfilled correspond to stars with anisotropic pressure. The solid (dashed) line correspond to a constant density neutron star with isotropic (anisotropic) pressure, while the star symbol is the black hole limit (discussed in Sec.~\ref{sec:entropy-area}). Observe that the entropies approach the black hole limit as the compactness increases in a way that is approximately independent of the equation of state.}
\label{fig:Sthermal}
\end{figure}

Figure~\ref{fig:Sthermal} shows the thermodynamic entropy scaled by $\bar{T}/M$ for a sequence of neutron stars with different compactnesses $C = M/R$, where $M$ is the mass of the star and $R$ is its radius. Each point in this figure requires its own solution to the differential equations of relativistic structure, as described in the previous paragraph. Observe that the thermodynamic entropy is approximately independent of the equation of state. Observe also that as the compactness increases, the thermodynamic entropy approaches a single point labeled with a star. This points corresponds to the black hole Hawking entropy, which we will discuss in Sec.~\ref{sec:entropy-area}. Of course, a sequence of neutron stars with isotropic pressure
can never exceed the so-called Buchdahl limit of $C=4/9$, and thus, the thermodynamic entropy sequences terminate there. For anisotropic stars, however, the sequence is allowed to continue, and the thermodynamic entropy approaches the black hole point much more closely, although it overshoots it by roughly $10\%$ as $C \to 1/2$ when using realistic equations of state. Observe also that for a constant density anisotropic star, the thermodynamic entropy approaches the black hole limit exactly, as we will also be showing analytically in Sec.~\ref{sec:entropy-area}.   
%

\subsection{Configurational Entropy}
\label{subsec:conf-entropy}

Another way to understand the concept of entropy is in terms of the number of arrangements possible for all the particles of a system, while at the same keeping some physical properties, such as energy, constant. Zurek and Thorne~\cite{Zurek:1985gd} showed that the entropy of a rotating and charged black hole can be understood as the logarithm of the number of configurations of the black hole as measured by distant observers, or roughly speaking, the amount of information it encloses. Building on these ideas, Gleiser \textit{et al}~\cite{Gleiser:2013mga,Gleiser:2015rwa} define the \emph{configurational entropy density} via
\be
\label{eq:conf-entropy-density}
s_{c} = - \int \hat{f}(\vec{k}) \ln\left[\hat{f}(\vec{k})\right] d^{3}k\,,
\ee
where $\hat{f}(\vec{k}) := f(\vec{k})/f_{\max}$, 
\be
f(\vec{k}) := \frac{|\tilde{\rho}(\vec{k})|^{2}}{\int |\tilde{\rho}(\vec{k})|^{2} d^{3}k}\,,
\ee
the overhead tilde stands for the Fourier transform
\ba
\label{eq:rho-tilde}
\tilde{\rho}(\vec{k}) &=& \frac{1}{(2 \pi)^3} \int \rho(r) \,e^{2 \pi i \vec{k} \cdot \vec{x}}\, d^{3}r \nn \\
&=&\frac{1}{2 \pi ^2 \,k} \int_0^R \rho(r)\, r\, \sin (k r) \,dr\,,
\ea
and $\rho(r)$ is the energy density profile. The quantity $f_{\max}$ is the maximum fraction, given by the system's longest physical mode with $k_{\min} = \pi/R$. The normalization of $\hat{f}(\vec{k})$ is critical to guarantee that $\hat{f} < 1$ and to ensure the argument of the natural logarithm is dimensionless. Although at first sight the definition of configurational entropy above seems arbitrary, it is inspired from Shannon's information entropy, and it has been shown to allow for stability bounds of astrophysical compact objects~\cite{Gleiser:2015rwa}.

For neutron stars, the configurational entropy can be calculated straightforwardly from the equations presented above. As in the previous subsection, we begin by choosing a central density and numerically solving the differential equations of relativistic structure for a specific equations of state. This solution yields the energy density profile, which we can then Fourier transform and use to calculate $\hat{f}(\vec{k})$. The latter, can then be used directly in the configurational entropy density of Eq.~\eqref{eq:conf-entropy-density}. As in the previous subsection, we then repeat this procedure for different central densities to find a neutron star sequence of varying compactness, in which we store the radius, total mass, compactness and configurational entropy for each member of the sequence. We conclude by repeating this calculation for different equations of state.

\begin{figure}[htb]
\includegraphics[width=0.48\textwidth]{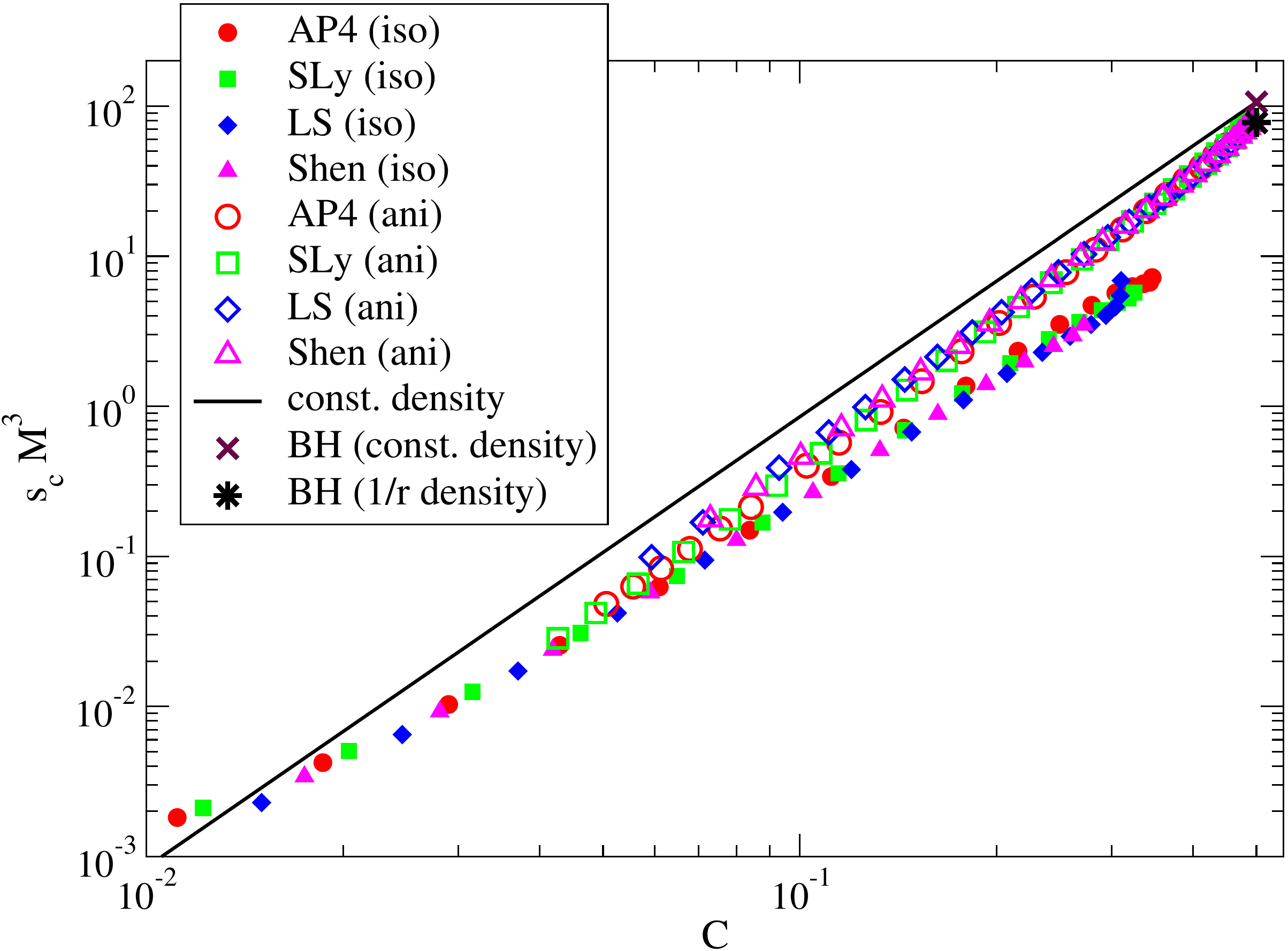}
\caption{Configurational entropy of a sequence of neutron stars of different compactness $C$, scaled by the mass cubed, for isotropic (filled symbols) and anisotropic (unfilled symbols) neutron stars. The solid line corresponds to a constant density star, while the star symbol is the black hole limit for a $1/r$ energy density profile. Observe that the entropies approach the black hole limit as the compactness increases in a manner that is approximately independent of the equation of state. 
 }
\label{fig:Sconfig}
\end{figure}

Figure~\ref{fig:Sconfig} shows the configurational entropy density scaled by the mass cubed for a sequence of isotropic and anisotropic neutron stars of varying compactness with the same equations of state considered in Sec.~\ref{subsect:thermo-entropy}. As in the case of the thermodynamic entropy, observe that the configurational entropy is approximately independent of the equation of state within the class of isotropic models and the class of anisotropic models. Moreover, as before, the configurational entropy approaches a point labeled with a star, which we will later identify with the black hole limit in Sec.~\ref{sec:entropy-area}. As in the previous subsection, the isotropic sequence terminates at a threshold compactness for each equation of state, but the anisotropic sequence continues to approach the black hole limit.

\begin{figure}[htb]
\includegraphics[width=0.48\textwidth]{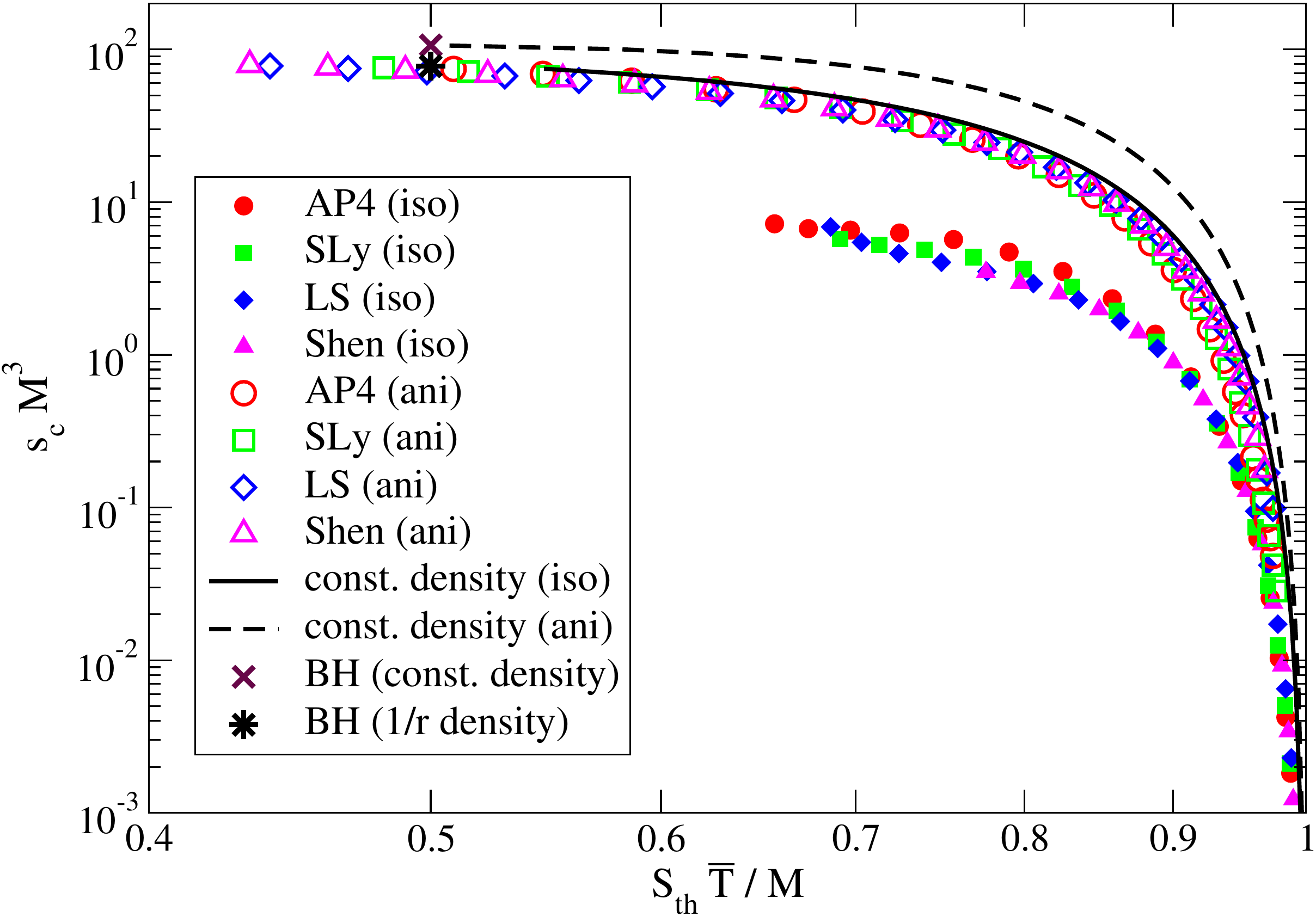}
\caption{Relation between the configurational and the thermodynamic entropies for a sequence of neutron stars of varying central density, using different realistic equations of state for isotropic and anisotropic stars. The meaning of each symbol is similar to that of Fig.~\ref{fig:Sconfig}. Observe that this relation is approximately equation of state independent within the class of isotropic and the class of anisotropic models. }
\label{fig:Sconfig-SThermal}
\end{figure}

With the configurational entropy as a function of compactness (Fig.~\ref{fig:Sthermal}) at hand, we can now invert it and combine it with the thermodynamic entropy as a function of compactness (Fig.~\ref{fig:Sconfig}) to find a relation between the thermodynamic and the configurational entropy. Figure~\ref{fig:Sconfig-SThermal} shows this relation for a sequence of isotropic and anisotropic neutron stars of varying central density, using the same equations of state as in Sec.~\ref{subsect:thermo-entropy}. Observe that, once more, the relation between these entropies is approximately independent of the equation of state. Observe also that as the thermodynamic entropy approaches the black hole Hawking entropy, the configurational entropy approaches the same configurational black hole limit discussed above. 
%

\section{Approximate Universality of the Entropy of Neutron Stars}
\label{sec:SILQ}

In this section, we study the degree of approximate universality (i.e. the approximate insensitivity with variations in the equation of state) of the entropy as a function of the compactness and as a function of the moment of inertia of neutron stars. The latter is  calculated by constructing a slowly-rotating neutron star valid to first order in its spin angular momentum~\cite{hartle1967,hartlethorne}. We begin by presenting a measure of universality using the numerical results obtained in the previous section. We conclude with analytical estimates of universality, obtained perturbatively in a leading-order expansion in small compactness. 

\subsection{Approximate Universal Relations}

Let us begin by studying the degree of variability of the entropy. Figures~\ref{fig:Sthermal} and~\ref{fig:Sconfig} have already presented the entropy-compactness ($S$--$C$) relations, and even by eye it is clear that these relations are approximately insensitive to the equation of state, i.e. approximately universal, within the isotropic and the anisotropic classes. Recently, two of us discovered that the degree of universality of certain global quantities can be enhanced when using the normalized moment of inertia $\bar{I} = I/M^{3}$, the normalized quadrupole moment or the Love number as the independent variable~\cite{I-Love-Q-Science,I-Love-Q-PRD,Yagi:2016bkt} (instead of the compactness in our case). Let us then study the normalized moment of inertia-entropy ($\bar{I}$--$S$) relation using the numerical results obtained in the previous section. 

\begin{figure*}[htb]
\includegraphics[width=0.495\textwidth,clip=true]{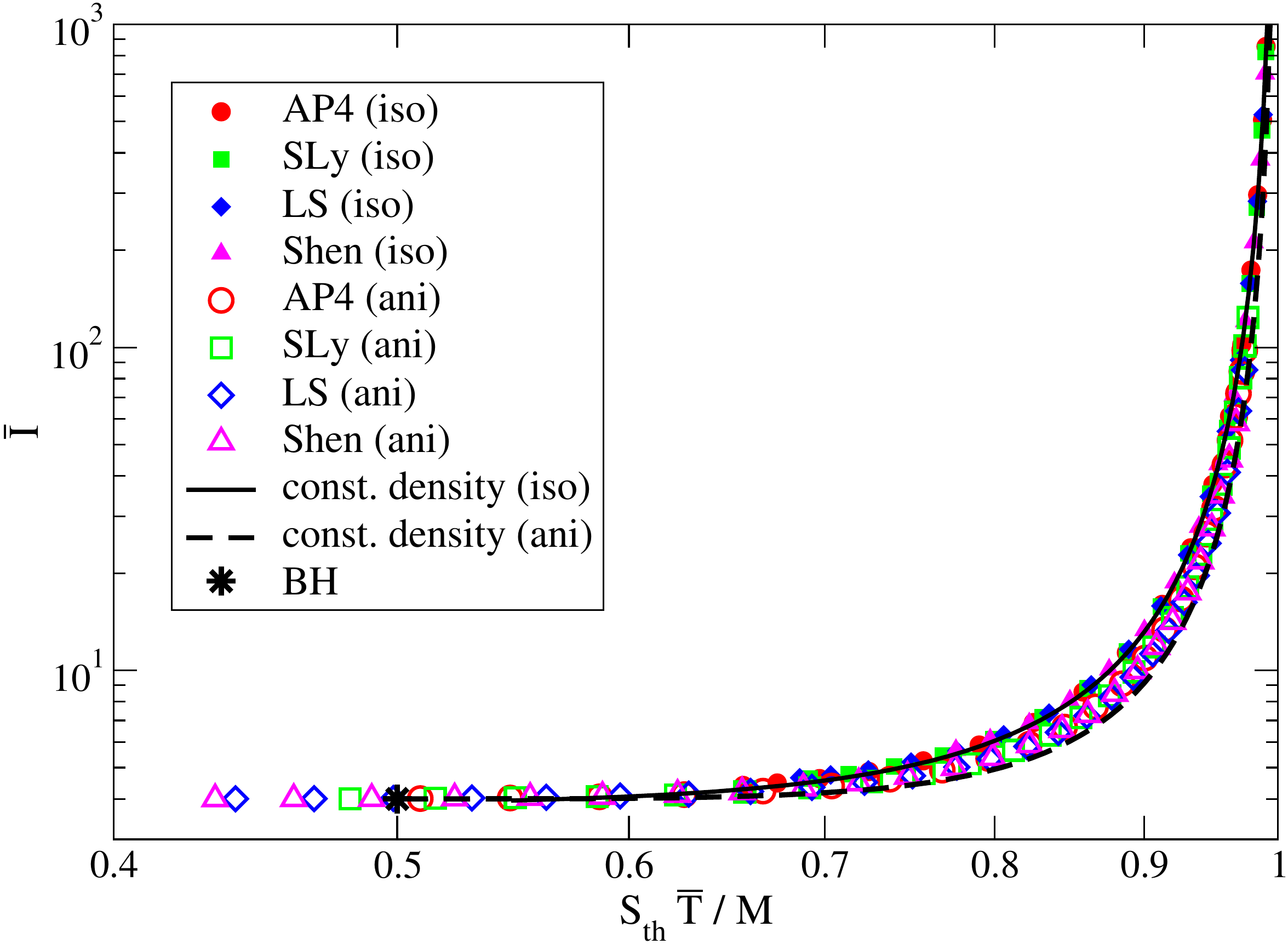}
\includegraphics[width=0.495\textwidth,clip=true]{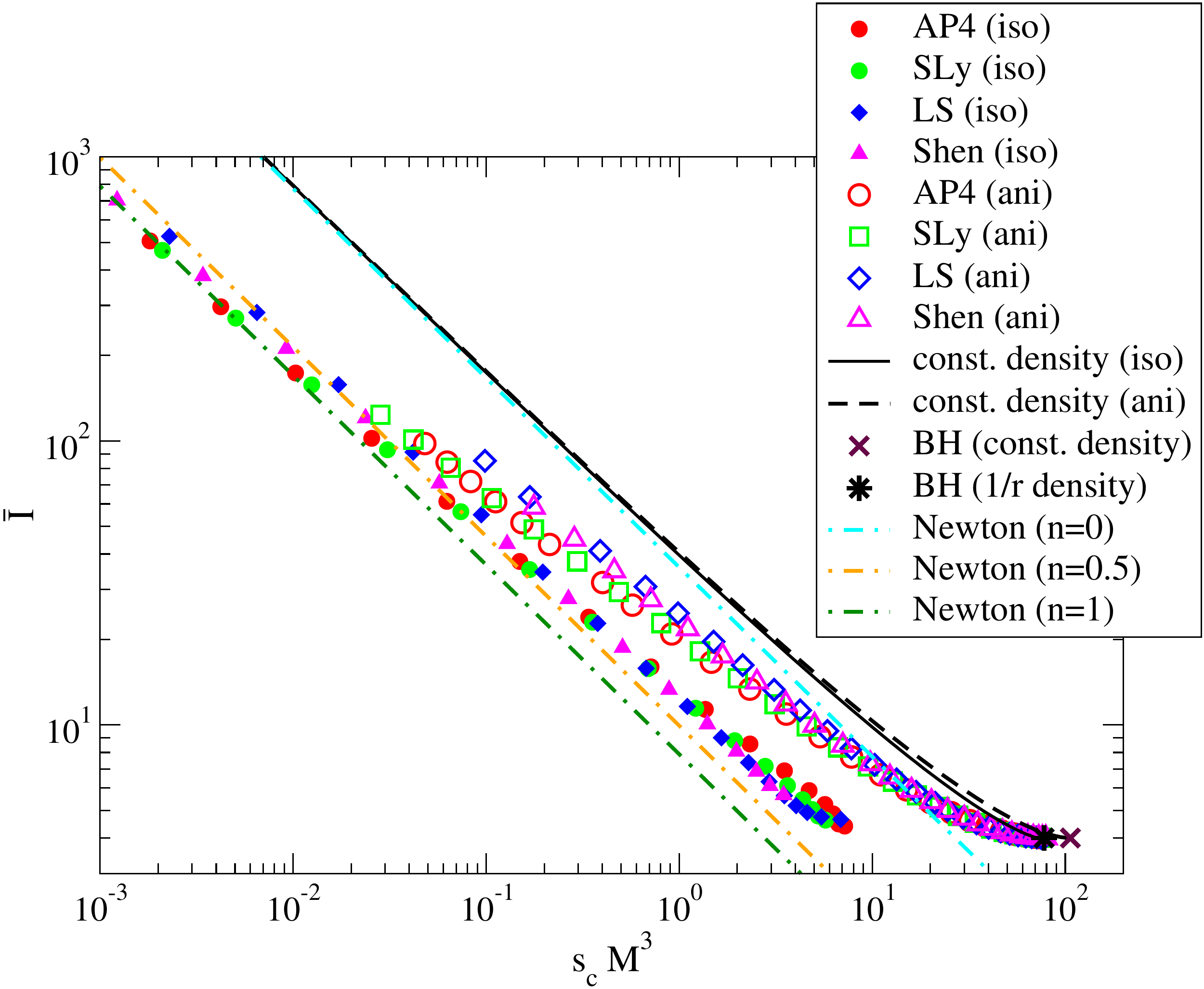}
\caption{Normalized moment of inertia $\bar I$ against the thermodynamic entropy (left-panel) and the configurational entropy (right panel) for a sequence of isotropic and anisotropic neutron stars with different equations of state. The values of the moment of inertia and the entropy for a black hole are shown with a star symbol on the left panel, and with a star and a cross symbol on the right panel for a constant density and a $1/r$ density profile respectively. Observe that the entropies approach the black hole result in the black hole limit in a manner that is approximately independent of the equation of state. For the configurational entropy case, we also present the relation for Newtonian polytropes.}
\label{fig:I-S}
\end{figure*}
Figure~\ref{fig:I-S} shows the $\bar{I}$--$S$ relation using the thermodynamic entropy scaled by the renormalized temperature and the mass (left panel) and using the configurational entropy scaled by the mass cubed (right panel). As in the $S$--$C$ case, there is a clear approximate universality in the $\bar{I}$--$S$ relation within the isotropic and the anisotropic classes. In the $\bar{I}$--$S$ thermodynamic case, however, both the isotropic and the anisotropic sequences seem to be approximately universal. Observe also that, as expected, the $\bar{I}$--$S$ relations also approach the black hole result as the sequence approaches the black hole limit; recall here that the normalized moment of inertia of a slowly-rotating Kerr black hole is $\bar{I} = 4$, as explained for example in~\cite{I-Love-Q-PRD,Yagi:2016bkt}.  

Let us now attempt to quantify better the degree of universality present in the $S$--$C$ and the $\bar{I}$--$S$ relations. In fact, to have a better comparison, we consider the compactness as a function of the entropy ($C$--$S$) instead of the $S$--$C$  relation. We will compare the relations for a given equation of state against averaged relations. We define the \emph{averaged} $C$--$S$ curve as follows:
\be
C_{\rm{avg}}(S) = \frac{1}{N} \sum_{i}^{N} C_{i}(S)\,,
\ee
where $N$ is the number of equations of state considered within each class and $C_{i}(S)$ is the interpolated $C$--$S$ relation for the $i$th  equation of state in the class. Similarly, we define the \emph{averaged} $\bar{I}$--$S$ curve as follows:
\be
\bar{I}_{\rm{avg}}(S) = \frac{1}{N} \sum_{i}^{N} \bar{I}_{i}(S)\,,
\ee
where again $N$ is the number of equations of state considered in each class, and $\bar{I}_{i}(S)$ is the interpolated $\bar{I}$--$S$ relation for the $i$th equation of state in the class. We will repeat this calculation for the thermodynamic entropy (scaled by the renormalized temperature and the mass) and for the configurational entropy (scaled by the mass cubed) to thus obtain 8 different averaged relations (2 relations, 2 different entropies, and 2 different classes). 

With the averaged relations calculated, we can then construct a measure of non-universality for the $i$th equation of state as follows
\begin{align}
(\bar{I}{\rm{-}}S \;\; {\rm{non}}{\rm{-}}{\rm{universality}})_i &= \frac{\left|\bar{I}_{i}(S) - \bar{I}_{\rm{avg}}(S)\right|}{\bar{I}_{i}(S)}\,,
\\
(C{\rm{-}}S  \;\; {\rm{non}}{\rm{-}}{\rm{universality}})_i &= \frac{\left| C_{i}(S) - C_{\rm{avg}}(S)\right|}{C_{i}(S)}\,.
\end{align}
If these non-universality measures are zero exactly, then the universality is exact and the relations are perfectly independent of the equation of state. If, on the other hand, the non-universality measures are small but non-zero, then we will say the universality is approximate, with the measures quantifying their degree.
 
\begin{figure*}[htb]
\includegraphics[width=0.495\textwidth,clip=true]{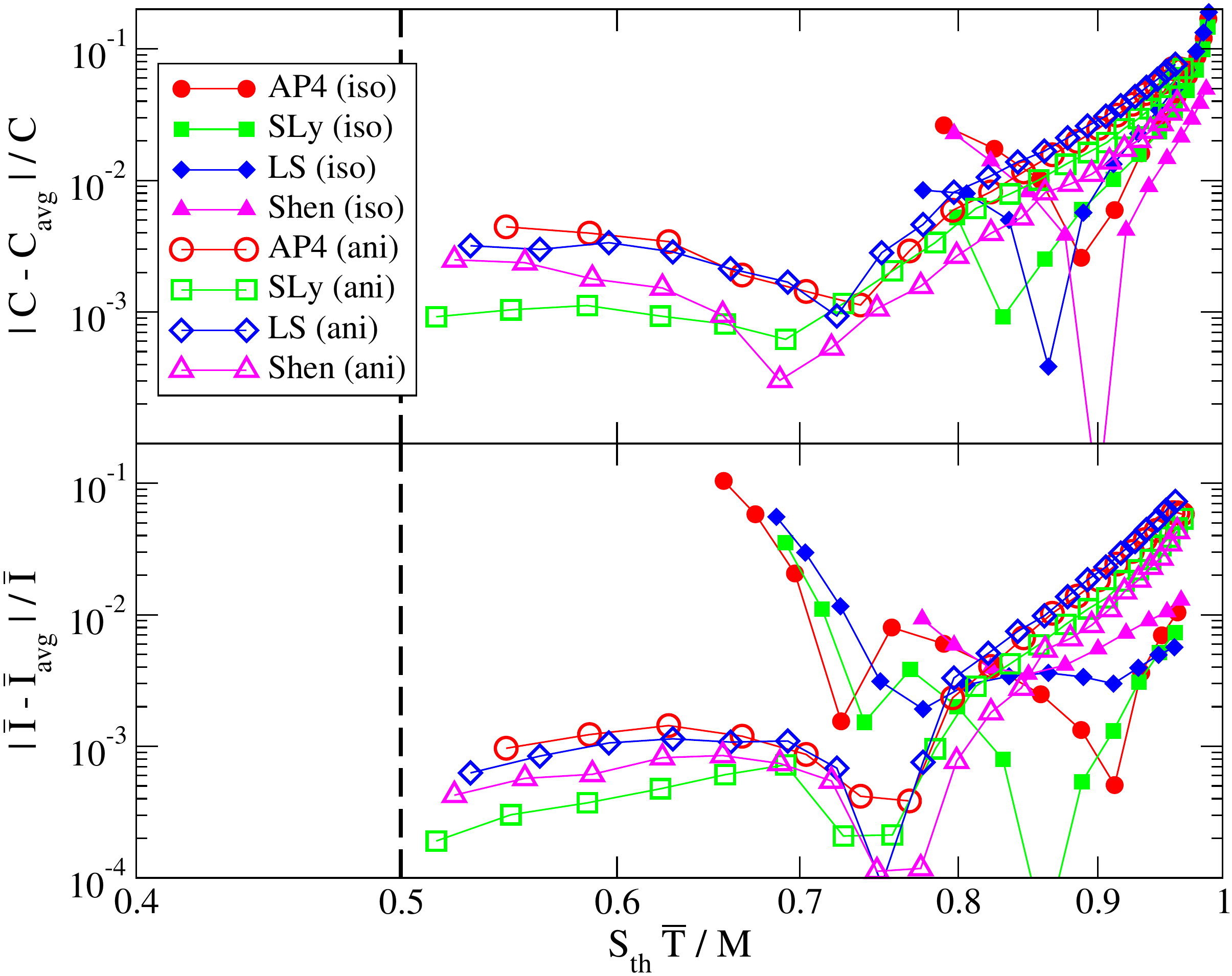}
\includegraphics[width=0.495\textwidth,clip=true]{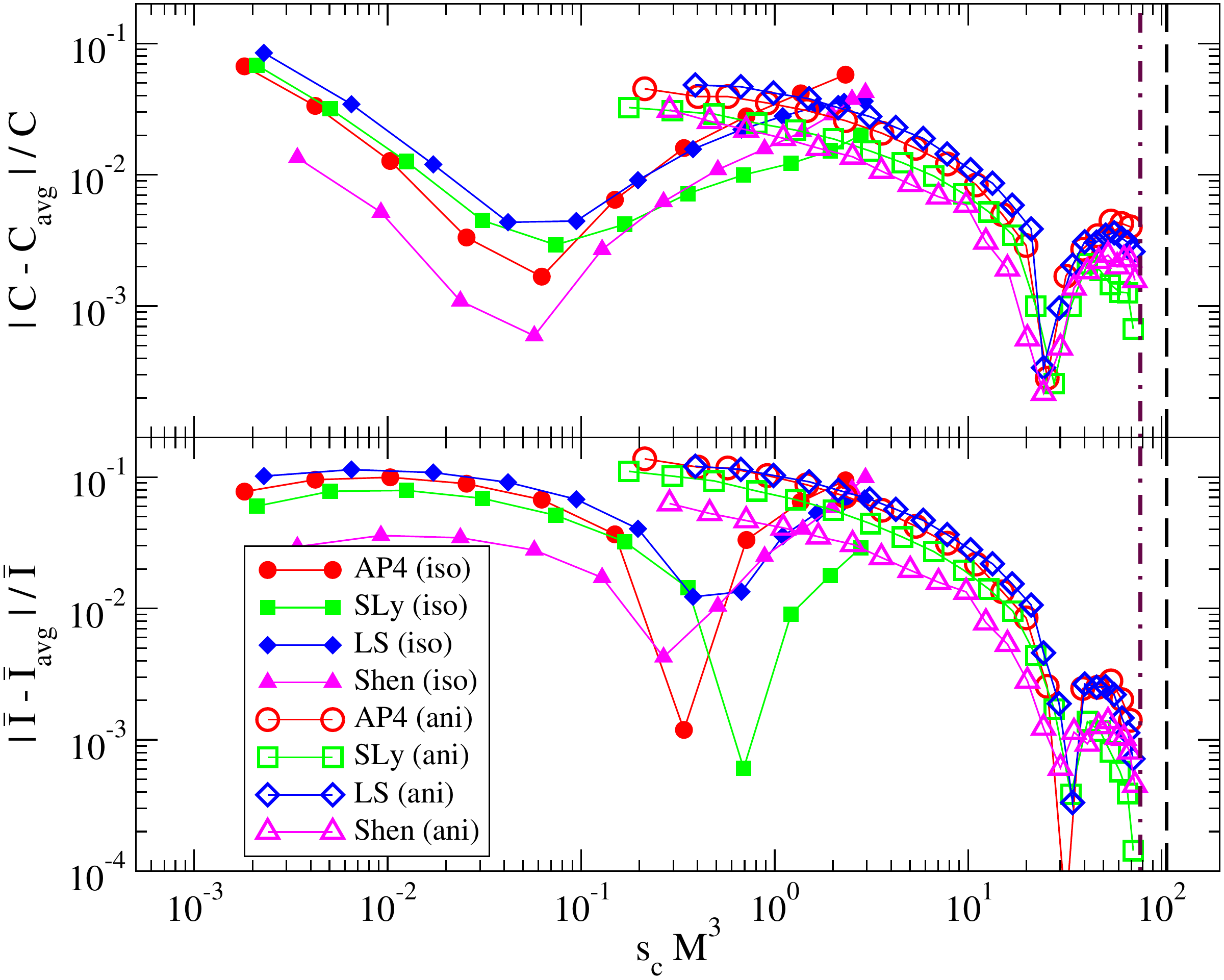}
\caption{Relative fractional difference in the $C$--$S$ (top panel) and the $\bar I$--$S$ (bottom panel) relations calculated between averaged relations and the relations obtained with a fixed equation of state. We present the thermodynamic (left panel) and configurational (right panel) entropies. Vertical lines show the black hole limits (for the configurational entropy, the black dashed one corresponds to the constant density profile, while the brown dotted-dashed one corresponds to the $1/r$ density profile). Observe that the variability of the relation, relative to the averaged ones, is less than $10\%$ in most cases, becoming more universal as the anisotropic sequence approaches the black hole limit.
}
\label{fig:Frac-Diff}
\end{figure*}

Figure~\ref{fig:Frac-Diff} presents the non-universality measures for the $C$--$S$ (top panels) and the $\bar{I}$--$S$ (bottom panels) relations for the scaled thermodynamic entropy (left panel) and the configurational entropy (right panel). Observe that the relations are approximately universal to better than 10\% in most cases, with no noticeable improvement in the universality between the $C$--$S$ and the $\bar{I}$--$S$ relations. Moreover, observe that as the anisotropic sequence approaches the black hole limit, the universality improves by over 2 orders of magnitude reaching variabilities as low as ${\cal{O}}(0.1\%)$. The loss of universality in the $(S_{\rm{th}} \bar{T}/M) \to 1$ limit is due to great increase in steepness in the $C$--$S_{\rm{th}}$ and $\bar{I}$--$S_{\rm{th}}$ relations shown in Figs.~\ref{fig:Sthermal} and~\ref{fig:I-S}. 
%

\subsection{Newtonian Limit}

Let us now analytically estimate the neutron star entropies in the Newtonian limit, and compare these estimates to our numerical calculations. The Newtonian limit corresponds to perturbatively expanding any physical quantity, like the entropy, in small compactness $C \ll 1$, and retaining only the leading-order-in-compactness result. In this sense, the Newtonian limit is analogous to a post-Minkowskian or weak-field limit.

The Newtonian calculation for thermodynamic entropy is simple. In the Newtonian limit, the quantities that enter the integrand of Eq.~\eqref{eq:therm-entropy-new} reduce to $g_{tt} \to 1$, $g_{rr} \to 1$ and $p \ll \rho$, which leads to
\be
S_\mrm{th}^{\mrm{Newt}} = \frac{4 \pi}{\bar{T}} \int_{0}^{R} r^{2} \rho \, dr + {\cal{O}}(C)= \frac{M}{\bar T} + {\cal{O}}(C)\,,
\ee
and thus $(S_\mrm{th}\, \bar T /M) \to 1$ in the Newtonian limit. Figure~\ref{fig:Sthermal} shows clearly that indeed $(S_\mrm{th}\, \bar T /M)$ approaches unity as $C \to 0$.

Deriving the Newtonian behavior of the configurational entropy is slightly more complicated. For simplicity, let us focus on  Newtonian polytropes with the equation of state
\be
p = K \rho^{1+1/n}\,,
\ee
where $K$ and $n$ are constants. Typical neutron star equations of state can be approximated by the above polytropic equation of state with a polytropic index $n \in (0.5,1)$. 

The configurational entropy for polytropes is given by~\cite{Gleiser:2015rwa}
\be
\label{eq:S}
s_c =  - 4 \pi \alpha^{-3} \int^\infty_{\kappa_\mathrm{min}} \tilde f(\kappa) \log \left[ \tilde f(\kappa) \right] \, \kappa^2 \, d\kappa\,,
\ee
where $\kappa = \alpha \, k $ and
\be
\alpha^2 \equiv \frac{K}{4\pi} (1+n) \rho_0^{-1+1/n}\,,
\ee
with $\rho_0$ the central density. We define $\kappa_\mathrm{min} \equiv \pi/\xi_R$ with $\xi_R$ representing the location of the stellar surface in the dimensionless radial coordinate $\xi = r/\alpha$. Equation~\eqref{eq:S} also makes use of $\tilde f (\kappa)$, which is defined as
\be
\label{eq:f}
\tilde f (\kappa) = \frac{h(\kappa)^2}{h(\kappa_\mathrm{min})^2}\,,
\ee
with\footnote{We drop the constant factor in $h$ since it cancels in Eq.~\eqref{eq:f}.}
\be
\label{eq:h}
h(\kappa) \equiv  \int^{\xi_R}_0 [\vartheta(\xi)]^{n} \,  \xi \, \sin (\kappa \xi) \, d\xi \,.
\ee
Here $\vartheta(\xi) = [\rho(r)/\rho_0]^{1/n}$ is the Lane-Emden function, a dimensionless measure of the energy density, defined as a solution to the Lane-Emden equation
\be
\frac{1}{\xi^2} \frac{\partial}{\partial \xi} \left( \xi^2 \frac{\partial}{\partial \xi} \right) \vartheta = - \vartheta^n\,,
\ee
with the boundary condition $\vartheta (0)=1$ and $\vartheta'(0)=0$.
Using further that
\be
\label{eq:alpha}
\alpha = \frac{R}{\xi_R}\,,
\ee
one finds
\be
\label{eq:sc-poly}
s_c \, M^3 =  - 4 \pi \, C^3\,\xi_R^{3} \int^\infty_{\kappa_\mathrm{min}} \tilde f(\kappa) \log \left[ \tilde f(\kappa) \right] \, \kappa^2 \, d\kappa\,.
\ee

The Lane-Emden equation can be solved only numerically for a generic polytropic index $n$, but one can find analytic solutions for certain polytropic indices, including $n=0$ (constant density stars) and $n=1$. For such a constant density star, $\vartheta^{(n=0)} = 1-\xi^2/6$ with $\xi_R^{(n=0)} = \sqrt{6}$, which then implies  
\be
\tilde f^{(n=0)} =\frac{\pi ^4 \left[\sin \left(\sqrt{6} \kappa \right)-\sqrt{6}
   \kappa  \cos \left(\sqrt{6} \kappa \right)\right]^2}{216 \,
   \kappa ^6}\,,
\ee
and one can perform the integration in Eq.~\eqref{eq:sc-poly} numerically to find 
\be
\label{eq:sc-n0}
s_c^{(n=0)} M^3 \approx 8.5 \times 10^2 \, C^3\,. 
\ee
For an $n=1$ polytrope, $\vartheta^{(n=1)} = \sin \xi/\xi$ and $\xi_R^{(n=1)} = \pi$, which then implies
\be
\tilde f^{(n=1)} =\frac{4 \sin ^2(\pi  \kappa )}{\pi ^2 \kappa^2 \left(1 -\kappa
   ^2\right)^2}\,,
\ee
and one finds numerically that
\be
\label{eq:sc-n1}
s_c^{(n=1)} M^3 \approx 1.7 \times 10^2 \,C^3\,.
\ee
We see then that the configurational entropy in the Newtonian limit is simply proportional to compactness cubed.

Let us now consider the Newtonian limit of the dimensionless moment of inertia $\bar I$. Our starting point is the integral expression for the moment of inertia to leading-order in the weak-field~\cite{hartle1967}
\be
I = \frac{8\pi}{3} \int_0^R \rho r^4 \, dr = \frac{8\pi}{3} \alpha^5\, \rho_0\, I_4\,,
\ee
where
\be
I_\ell \equiv \int_0^{\xi_R} \vartheta^n \, \xi^\ell \, d\xi\,.
\ee
Using that the weak-field expression for the total mass is
\be
M = 4\pi \int_0^R \rho\, r^2 \, dr = 4\pi \, \alpha^3\,\rho_0\,  I_2\,,
\ee
one finds
\be
\bar I \equiv \frac{I}{M^3} = \frac{2}{3} \frac{1}{\xi_R^2} \frac{I_4}{I_2} \frac{1}{C^2} + {\cal{O}}(C^{-1})\,.
\ee
For $n=0$ and $n=1$ polytropes, $\bar I$ is then given by
\be
\label{eq:Ibar-poly}
\bar I^{(n=0)} = \frac{2}{5} C^{-2}\,, \quad \bar I^{(n=1)} = \left( \frac{2}{3} - \frac{4}{\pi^2} \right) C^{-2}\,.
\ee

We can now combine these two sets of results to find the $\bar{I}$--$S$ relation for the configurational entropy. Solving for the compactness in Eqs.~\eqref{eq:sc-n0} and~\eqref{eq:sc-n1} in terms of $s_c$ and inserting this in Eq.~\eqref{eq:Ibar-poly}, we find 
\begin{align}
\bar I^{(n=0)} &\approx 36  \left(s_c^{(n=0)} M^3\right)^{-2/3}\,,
\\
\bar I^{(n=1)} &\approx 7.9  \left(s_c^{(n=1)} M^3\right)^{-2/3} \,.
\end{align}
In the right panel of Fig.~\ref{fig:I-S}, we present the above Newtonian relations for $\bar I$ against $s_c M^3$ for $n=0$ and $n=1$ polytropes. For reference, we also show the relation obtained numerically for an $n=0.5$ polytrope. Observe that the relations for neutron stars approach either the $n=0.5$ or $n=1$ relation for small $s_c M^3$, which corresponds to the weak-field regime. Observe also that the $n=0.5$ relation is much closer to the $n=1$ relation than the $n=0$ one is. This (semi-)analytically explains the universality in the relations found here, since neutron star equations of state can roughly be bracketed by polytropes with index between $n=0.5$ and 1.

\section{Comparisons with Black Hole Entropy}
\label{sec:entropy-area}

In this section, we study how the rescaled thermodynamic entropy and the rescaled configurational entropy of a sequence of neutron stars behaves as the sequence approaches the compactness of a black hole. We begin by comparing the latter to the entropy of a Schwarzschild black hole. We conclude with a discussion of how the sequence approaches the entropy-area law of black holes.   

\subsection{Approach to the Black Hole Entropy}

For black holes, Bekenstein~\cite{Bekenstein:1973ur,Bekenstein:1974ax} and Hawking~\cite{Hawking:1974sw,Hawking:1974rv} showed that the thermodynamic entropy scales with the horizon surface area, namely
\be
\label{eq:BH-entropy}
S_{\BH} = \frac{k_{B}}{4} \frac{A}{\ell_{p}^{2}}\,,
\ee
where $\ell_{p} = \sqrt{\hbar}$ is the Planck length and $k_{B}$ is the Boltzmann constant. Using that the Hawking temperature is
\be
T_{\HH} = \frac{\ell_{p}^{2}}{8 \pi M k_{B}}\,,
\ee
one can then write 
\be
S_{\BH} = \frac{1}{2} \frac{M}{T_{\HH}}\,,
\ee
where we have used the surface area of a Schwarzschild black hole $A = 4 \pi R_{\BH}^{2} = 16 \pi M^{2}$. We can go one step further and construct the following dimensionless number by combining the results presented above
\be
\label{eq:STM}
S_{\BH} \left(\frac{T_{\HH}}{M}\right) = \frac{1}{2}\,.
\ee
Note that for a black hole, the chemical potential vanishes and thus $\bar T$ in Eq.~\eqref{eq:Tbar} is equivalent to $T_\HH$. 

Before proceeding, let us make some quick order of magnitude estimates. The Bekenstein-Hawking entropy for a $10 M_{\odot}$ black hole is $S_{\BH} = 10^{79} k_{B} = 10^{12} \; {\rm{m}}/{\rm{K}}$ in geometric units, because $T_{\BH} = 6 \times 10^{-9} \; {\rm{K}}$. The thermodynamic entropy of a $1.4 M_{\odot}$ neutron star with a radius of $R_{\NS} \approx 11 \; {\rm{km}}$ is $S_{\NS} \approx M_{\NS}^{2}/(R_{\NS} T_{\NS)} \approx 4 \times 10^{-4} \; {\rm{m}}/{\rm{K}}$, which is much smaller than that of a black hole, essentially because the neutron star temperature $T_{\NS} \approx 10^{6} \; {\rm{K}} \gg T_{\BH}$.  Yet still, the neutron star rescaled entropy $S_{\NS} (T_{\NS}/M_{\NS}) \approx 0.19$ is quite close to the black hole rescaled entropy $S_{\BH} (T_{\BH}/M) = 1/2$.

The configurational entropy for black holes is not actually well-defined formally. This is simply because black holes are vacuum solutions with $\rho=0$ except at the singularity, a point that is typically removed from the manifold. If one sets $\rho=0$ in the configurational entropy equations of Sec.~\ref{subsec:conf-entropy}, then the latter vanishes. One can instead construct an effective energy density for a black hole, as done in~\cite{Braga:2016wzx}, for example arguing that the Fourier transform of the density must be square-integrable and the density must be singular at the singularity. Doing so, one finds that $\rho = 2 M/(R_{\BH}^{2} r)$ and numerically that $s_{c} \approx 77.7/M^{3}$. However, one should note that this choice of density functional is not necessarily unique, and thus, that there is not a clear way to relate the configurational entropy to the Hawking entropy. 

Now that we have discussed the thermodynamic and configurational entropies for black holes, let us consider the limit of these entropies for neutron stars as we approach the black hole limit. We are here focusing on non-rotating stars in Schwarzschild like coordinates, so the black hole limit is approached as the radius of the neutron star approaches the event horizon of a Schwarzschild black hole $R \to 2 M$, and thus, as the compactness $C = M/R \to 1/2$. Figures~\ref{fig:Sthermal} and~\ref{fig:Sconfig} have already presented the rescaled thermodynamic and configurational entropies as a function of compactness. Referring back to these figures, we see that the rescaled thermodynamic entropy approaches the rescaled Hawking entropy as $C \to 1/2$ for anisotropic stars (recall that for isotropic stars, the Buchdahl limit prevents us from taking the black hole limit). In fact, for a constant density anisotropic star, the rescaled thermodynamic entropy goes exactly to the rescaled Hawking entropy in the black hole limit, a result we prove analytically in Sec.~\ref{subsec:scaling}. For the same sequence, the rescaled configurational entropy approaches a value close but slightly above the rescaled configurational entropy of a black hole, but as we discussed in the previous paragraph, the latter is not unique due to its intrinsic dependence on the energy density.    

Let us stress that the universal relations and the approach to the black hole limit only occurs for an appropriate \emph{rescaled entropy} and not for the entropies themselves. In fact, the Bekenstein-Hawking entropy is obviously much larger than the thermodynamic entropy of a typical neutron star by many orders of magnitude, and thus there is a large gap between the two. However, such a gap seems to disappear once one renormalizes the entropies using the stellar mass and the temperature. It is intriguing that although the entropy is vastly different between neutron stars and black holes, the universality in the rescaled entropy relations holds throughout the neutron star sequence and all the way up to the black hole limit. 

The results discussed above are not the first sign that the thermodynamic entropy of neutron stars approaches the black hole limit as the compactness increases. Pretorius, Vollick and Israel~\cite{Pretorius:1997wr} have shown that as a thin spherical shell contracts from spatial infinity down to its horizon, the thermodynamic entropy approaches one quarter of its area. This result was followed up by Oppenheim, who studied the thermodynamic entropy of a self-gravitating object composed of a series of thin shells of arbitrary composition~\cite{Oppenheim:2001az}, as well as the thermodynamic entropy of a gravitating perfect fluid of constant density~\cite{Oppenheim:2002kx}. He found that when gravitational effects are ignored, the thermodynamic entropy scales with the volume of the star, while when they are included, the thermodynamic entropy scales with the area in the limit as the boundary of the system approaches its would-be event horizon. This area scaling is what one would expect of the entropy of black holes, as shown in Eq.~\eqref{eq:BH-entropy}. The results presented above complement and extend these previous findings, by considering gravitating perfect fluids with more general equations of state and showing that these different neutron star sequences approach the black hole limit in an approximately universal manner. 

\subsection{Approach to the Black Hole Entropy-Area Law} 
\label{subsec:scaling}

Let us now attempt to understand whether the thermodynamic and configurational entropies scale with the area of the compact object in the black hole limit.  Following~\cite{Oppenheim:2002kx}, we begin by rewriting the thermodynamic entropy as 
\be
\label{eq:Sth_in_gamma}
S_\mrm{th} = \gamma \beta M\,,
\ee
where we have defined the dimensionless constant 
\be
\gamma \equiv \frac{S_\mrm{th} \bar T}{M}\,,
\ee
and the inverse temperature $\beta \equiv 1/\bar T$. In the black hole limit, when pressure can be ignored, the first law of thermodynamics states that 
\be
\label{eq:first_law}
dM = \frac{1}{\beta} dS_\mrm{th}\,.
\ee
Taking the derivative of Eq.~\eqref{eq:Sth_in_gamma} and using Eq.~\eqref{eq:first_law}, one finds
\be
\frac{1-\gamma}{\gamma} \frac{dM}{M} = \frac{d \beta}{\beta}\,,
\ee
which can be integrated to yield
\be
\beta \propto M^{(1-\gamma)/\gamma}\,.
\ee
Inserting this back into Eq.~\eqref{eq:Sth_in_gamma}, one finds that in the black hole limit
\be
S_\mrm{th} \propto M^{1/\gamma}\,.
\ee
Thus, the quantity $\gamma$, which is actually what is plotted in Fig.~\ref{fig:Sthermal}, controls the inverse of the scaling index of $S_\mrm{th}$ with $M$. Thus, if one wishes to recover the scaling $S_\mrm{th}  \propto M^{2}$ in the black hole limit, then one must have that $\gamma \to 1/2$ as $C \to 1/2$, in agreement with Eq.~\eqref{eq:STM}. This is \emph{exactly} the behavior observed in Fig.~\ref{fig:Sthermal} independent of the equation of state.

Given the importance of the quantity $\gamma$ in the black hole limit, let us now derive it analytically for constant density stars. We could use directly the results of Oppenheim~\cite{Oppenheim:2002kx}, who calculated $\gamma$ for constant density, \emph{isotropic} stars (shown also in Fig.~\ref{fig:Sthermal}), but these stars can never reach the black hole limit due to the Buchdahl limit. As before, we overcome this problem by considering constant density stars with anisotropic pressure, thus extending another analysis of Oppenheim~\cite{Oppenheim:2001az}, who calculated $\gamma$ for spherical shells of matter with anisotropic pressure. Below, we will consider constant density, anisotropic \emph{stars} (not shells) and prove analytically that $\gamma \to 1/2$ as $C \to 1/2$. 

The anisotropic BL model has been studied in detail in the past, and for constant density stars, one can solve the equations of relativistic structure exactly. When $\lambda_\mrm{BL} = -2\pi$, one finds~\cite{1974ApJ...188..657B,Yagi:2016ejg}: 
\ba
\rho &=& \frac{3}{4\pi} \frac{C}{R^2}\,, \\
p&=& \frac{2\pi}{3} g_{rr} \rho^2 r^2\,, \\
g_{tt} &=& -\frac{(1-2 C)^{3/2}}{\sqrt{1- 2 C
   r^2/R^2}}\,, \\
g_{rr} &=& \left(1 - 2C\frac{r^2}{R^2}  \right)^{-1}\,, 
\ea
with $p$ representing the tangential pressure and the radial pressure vanishes. Substituting these expressions into Eq.~\eqref{eq:therm-entropy-new}, one finds
\begin{equation}
\gamma= \frac{3 C^2-3
   C+1+(2 C-1) \,
   _2F_1\left(-\frac{1}{4},1;\frac{1}{2};2 C\right)}{2 C^2}\,, 
\end{equation}
where $_{2}F_{1}$ is a hypergeometric function. One can clearly see that taking the $C \to 1/2$ limit, then $\gamma \to 1/2$ exactly, which then implies that $S_\mrm{th} \sim M^{2}$ exactly in this case.

Let us now discuss the configurational entropy in the black hole limit. For constant density stars, Eq.~\eqref{eq:rho-tilde} becomes
\be
\tilde{\rho}(\vec{k}) = \rho \frac{\sin (k R)-k R \cos (k R)}{2
   \pi ^2 k^3}\,,
\ee
and $\hat{f}(\vec{k})$ is then
\be
\hat{f}(\vec{k}) = \frac{\pi ^4 [\sin (k R)-k R \cos (k R)]^2}{k^6 R^6}\,.
\ee
Substituting this into Eq.~\eqref{eq:conf-entropy-density}, one can numerically calculate the integral to find
\be
s_c M^3 = 8.5 \times 10^2\, C^3\,,
\ee
which is shown in Fig.~\ref{fig:Sconfig} as a solid black line. Notice that this expression is exactly the same as the Newtonian relation in Eq.~\eqref{eq:sc-n0}. In the black hole limit, the configurational entropy for these stars $s_c M^3 \to 1.1\times 10^2$. This value is close but different from the value of the configurational entropy in the black hole limit when one uses a $1/r$ density profile motivated from~\cite{Braga:2016wzx}, which is $s_{c} M^{3} \to 77.7$ as already mentioned earlier. This demonstrates clearly that the configuration entropy in the black hole limit is sensitive to the particular choice of density profile. 


\section{Musings and Future Directions}
\label{sec:conclusions}

We have explored the connection between neutron stars and black holes as one approaches the threshold of gravitational collapse of neutron stars. In particular, we have used the entropy as a tool to assess how this threshold is approached by an equilibrium sequence of cold and isolated neutron stars with different, yet realistic, equations of state. We have found that certain rescaled entropy measures (a rescaled thermodynamic one and a rescaled information-theoretic one) satisfy approximately universal relations, with the universality increasing as one approaches the threshold of collapse. Moreover, we have found that a rescaled entropy approaches that of a black hole in a manner that is increasingly independent of the star's equation of state. Finally, we have found that the entropy of neutron stars scales more and more with the neutron star's surface area as one approaches the threshold of collapse, again in a manner that is increasingly independent of the star's internal structure. 

But why is this? The analytic and numerical results we have obtained do not provide a concrete answer to this very interesting question, but they do provide some hints and allow us to rule out some hypothesis. The simplest hypothesis one may consider is that the level of universality observed is related to how similar the equations of state of neutrons stars are. Indeed, all equations of state of neutron stars agree near the crust, but they can also greatly disagree above nuclear saturation density, which is reached very rapidly beyond the crust. This is precisely why different equations of state predict very different mass-radius relations for neutron stars. In spite of these great dissimilarities in their mass-radius curves, the rescaled entropy of neutron stars does not seem to care much about the equation of state, especially near the threshold of gravitational collapse.   

The rescaled entropy is not the only macroscopic quantity that is insensitive to variations in the neutron star equation of state. Previously, two of us found that the moment of inertia, the Love number and the rotation-induced quadrupole moment satisfy similar universal relations (the I-Love-Q relations)~\cite{I-Love-Q-Science,I-Love-Q-PRD}. A hypothesis for the existence of these relations that has not been ruled out is that of an emergent symmetry~\cite{Yagi:2014qua}: as one considers ever more compact stars, the lowest energy configuration is one in which isodensity contours are approximately self-similar ellipsoids. Perhaps, this same emergent symmetry may also be responsible for the universality in the entropy that we have observed, an idea that deserves further study. Such an idea is reminiscent of explanations of critical phenomena through the emergence of scale invariance, but the scales involved here are hugely dissimilar. 

Another idea one could explore in the future is whether the universality we have found is robust to the inclusion of strong magnetic fields or rapid rotation. In the I-Love-Q case, the universality was found to be robust to the latter for a fixed dimensionless spin angular momentum~\cite{Pappas:2013naa,Chakrabarti:2013tca,Yagi:2014bxa}, but not to the former~\cite{I-Love-Q-B}, thus breaking in proto-neutron stars (see~\cite{Martinon:2014uua,Bretz:2015rna} for related work) and in magnetars (but not in millisecond pulsars). One may expect the same to be true for the entropy of neutron stars, but this should be verified through numerical calculations. 

Perhaps the ultimate question one may wish to address is whether the results we have found have other implications to our understanding of strong gravity. In the past, the connection between black holes and thermodynamics has been used to argue that one could derive the Einstein equations entirely from thermodynamic principles~\cite{Jacobson:1995ab,Verlinde:2010hp}. It is intriguing to think of the connections between our results and this derivation near the threshold of gravitational collapse. This, and other connections to holography and thermodynamics, could also be explored in the future.    

\acknowledgments
We would like to thank Roberto Emparan, Marcelo Gleiser, Ted Jacobson, Harvey Reall, Damian Sowinski and Nikolaos Stergioulas for very useful discussions and comments.  
KY acknowledges support from NSF Award PHY-1806776. 
K.Y. would like to also acknowledge networking support by the COST Action GWverse CA16104.
NY acknowledges support from NASA EPSCoR award 80NSSC17M0041 and NSF grant W7386. 

\appendix
\section{Chemical Potential Relation}
\label{funky-relation}

In this appendix, we prove the following relation: 
\be
\label{eq:app-start}
\frac{\rho+p}{\rho_{b}} = \left(\frac{g_{tt}(R)}{g_{tt}}\right)^{1/2}\,,
\ee
which leads to Eq.~\eqref{eq:funky-relation} by using $n = \rho_b \, m_N$ with $m_N$ representing the mass of a nucleon and $\mu(R)$ is constant.
Taking the derivative of this relation, we find
\be
\label{eq:drho}
d\left(\frac{\rho}{\rho_{b}}\right) + d\left(\frac{p}{\rho_{b}}\right) = - \frac{1}{2} \left[g_{tt}(R)\right]^{1/2}  \frac{dg_{tt}}{g_{tt}^{3/2}}\,.
\ee
Let us now use the first law of thermodynamics for equations of state at zero-temperature
\be
d \left(\frac{\rho}{\rho_{b}}\right) = -p\, d \left(\frac{1}{\rho_{b}}\right)\,.
\ee
Finite temperate effects will affect this relation, but we assume they will produce small corrections. Using this relation, Eq.~\eqref{eq:drho} becomes
\be
\frac{1}{\rho_{b}} \frac{dp}{dr} = - \frac{1}{2} \left[ \frac{g_{tt}(R)}{g_{tt}}\right]^{1/2} \frac{1}{g_{tt}} \frac{d g_{tt}}{dr}\,.
\ee

Let us now use the Einstein equations for a perfect-fluid source to simplify the above relation. One fo the Einstein equations reads
\be
\frac{d g_{tt}}{dr} = -2 g_{tt} \frac{4 \pi r^{3} p + m}{r (r - 2 m)}\,,
\ee
where $m$ is a metric function defined via $g_{rr} = (1 - 2 m/r)^{-1}$. Using this relation, we then find that 
\be
\frac{1}{\rho_{b}} \frac{dp}{dr} = \left[ \frac{g_{tt}(R)}{g_{tt}}\right]^{1/2} \frac{4 \pi r^{3} p + m}{r (r - 2m)} \,.
\ee
Using Eq.~\eqref{eq:app-start} one more time, the above equation becomes
\be
\frac{dp}{dr} = \frac{ (\rho + p)(4 \pi r^{3} p + m)}{r (r - 2 m)} \,.
\ee
One recognizes the above differential equation as the Tolman-Volkhoff-Oppenheimer equation, obtained by combining the Einstein equations with the stress-energy tensor conservation equation. Therefore, it follows that Eq.~\eqref{eq:app-start} is satisfied provided the Einstein equations are satisfied, the stress-energy tensor conservation equation is satisfied and the first law of thermodynamics at zero-temperature is satisfied. Corrections to the latter due to finite temperate effects will lead to modifications of ${\cal{O}}(T)$ to Eq.~\eqref{eq:app-start}. 

\bibliography{master}
\end{document}